# Fairer Chess: A Reversal of Two Opening Moves in Chess Creates Balance Between White and Black

Steven J. Brams
*Department of Politics*
New York University
New York, USA
steven.brams@nyu.edu

Mehmet S. Ismail
*Department of Political Economy*
King's College London
London, UK
mehmet.s.ismail@gmail.com

*Abstract*—Unlike tic-tac-toe or checkers, in which optimal play leads to a draw, it is not known whether optimal play in chess ends in a win for White, a win for Black, or a draw.  But after White moves first in chess, if Black has a double move followed by a double move of White and then alternating play, play is more balanced because White does not always tie or lead in moves. Symbolically, *Balanced Alternation* gives the following move sequence: After White's (W) initial move, first Black (B) and then White each have two moves in a row (BBWW), followed by the alternating sequence, beginning with W, which altogether can be written as WB/<u>BW</u>/WB/WB/WB… (the slashes separate alternating pairs of moves). Except for reversal of the 3rd and 4th moves from WB to BW (underscored), this is the standard chess sequence.  Because Balanced Alternation lies between the standard sequence, which favors White, and a comparable sequence that favors Black, it is highly likely to produce a draw with optimal play, rendering chess fairer.  This conclusion is supported by a computer analysis of chess openings and how they would play out under Balanced Alternation.

*Keywords—combinatorial games, chess, fairness, sequencing*

## I. Introduction

The rules of chess have evolved over the past 1,500 years, but beginning in the 19th century they were standardized to facilitate national and international competition.  While the most sophisticated chess-playing computer programs are now able to defeat the best human players, we still seem no closer to answering the question of whether chess is *fair*: When the players make optimal choices, neither White, who moves first, nor Black, who moves second, can force a win, rendering the outcome a draw.

Although most chess experts believe that a draw is the product of optimal play (for different views, see https://en.wikipedia.org/wiki/First-move_advantage_in_chess), there is no proof of this.  Because of the astronomical number of choices in chess, no brute-force check of all possible moves, even with the fastest computers, can verify this.  Accordingly, we take an indirect approach by showing that a small variation in the standard rules of chess—giving Black and White double moves after White's initial move, which reverses the order of play of the 3rd and 4th moves from White-Black to Black-White—balances the opportunities of White and Black to win with optimal play.  If neither player has a decisive advantage, it is highly unlikely that either player can force a win for itself (and loss for its opponent), making the outcome of optimal play a draw.

By *optimal play*, we mean that (i) if one player defeats its opponent, its opponent could not have done better—drawn or won—by making different moves; (ii) if the outcome is a draw, neither player could have won by making different moves.

If both these conditions are met, the optimal strategies of the players constitute a Nash equilibrium in a 2-person zero-sum game of perfect information (like chess), so neither player would depart from its optimal strategy (if known). Zermelo (1913) showed in chess that either (i) one player can force a win or (ii) both players can force a draw with optimal play (for a translation of this paper and discussion of its results, see [5].

In principle, backward induction, by working backwards from a final move when checkmate or a rule that forces a draw (e.g., the three-fold repetition of moves) occurs, can be used to find optimal moves of the players in a finite game such as chess. In checkers, this approach was applied by [4] to prove, using multiple computers making calculations over almost two decades, that optimal play always ends in a draw.  But because chess is a far more complex game than checkers, a comparable calculation to ascertain the outcome of optimal play of chess appears beyond the capability of computers for the foreseeable future.

Accordingly, we take a different approach by making a sequencing argument for the variation in chess we mentioned earlier—that it gives Black more opportunity to win and thereby creates more of a balance between Black and White.  It also has a major practical advantage, obviating the need of contestants to play both Black and White in a tournament, for reasons we give in the concluding section.  In section 4, we supplement our sequencing argument with an analysis of four well-known first moves by White in standard chess and suggest how they would play out under Balanced Alternation.  In section 5, we conclude that Balanced Alternation would render chess fairer by creating more even-handedness between Black and White, which in section 2 we show is decidedly biased in favor of White.

## II. Statistics on Chess

In tournament games that have a winner in chess (*decisive games*), White on average beats Black in 55 percent of them (Elo ratings of 2100 or above), but for elite players (Elo ratings of 2700 or above), the winning percentage is 64 percent.  The proportion of draws also increases with skill from 35 percent for nonelite players to 58 percent for elite players (Adorján, 2004, p. 68). These statistics have not changed much in recent years: White on average enjoys about a 2:1 advantage over Black at the highest (human) level of play, but at this level most games end in a draw. (These statistics are based on chess games played under classical time controls. For more information, see https://www.chessgames.com/chessstats.html.)

González-Díaz and Palacios-Huerta (2016) report White's advantage in all expert matches (players with an ELO rating above 2500) between 1970 and 2010. They found the winning percentage of White to be 64%. Moreover, although each player plays an equal number of games as White and Black in a match, they found that the winning percentage of the player who plays the first game as White to be 57%, which rises to 62% when only elite players (with an ELO rating above 2600) are considered. These findings illustrate the advantage enjoyed by White when each player plays an equal number of games as White and Black.

Statistics from computer play of chess by the strongest programs substantially amplify the advantage of White at the same time that they increase the proportion of draws. In 2020, when the reputedly most powerful chess engine in the world, AlphaZero, played against itself in 10,000 games, taking one minute per move, White won in 86 percent of the decisive games, but these games constituted only 2 percent of the total—98 percent were draws [6].

Other expert programs, including Leela Chess Zero and Stockfish, when pitted against each other in the superfinal of the unofficial world computer chess championship (TCEC), give White even greater odds of winning, but the outcome is still a draw in the large majority of games (see https://tcec-chess.com). Despite the fact that computer programs start play from 50 preselected opening positions in the TCEC superfinal (once as White and once as Black), it is remarkable that Black has not won a single game in the last two TCEC superfinals. All 49 decisive games were won by White, which was either Stockfish or Leela Chess Zero in the Superfinal tournament.

Although the forgoing statistics indicate that chess is biased against Black in decisive games, Black is usually able to survive by drawing. It is, nevertheless, surprising that White, by making the first move, is able to achieve almost a 6:1 advantage of winning in decisive games, based on the aforementioned AlphaZero statistics. Whether White has an *inherent* advantage—can force a win when both players make optimal choices—remains an open question.

More light would be shed on this question if the two leading machine-learning chess programs, AlphaZero and Leela Chess Zero, were taught to play with our proposed change in the order of the 3$^{rd}$ and 4$^{th}$ moves from White-Black to Black-White. Still, even if this change enabled Black to win a greater proportion of decisive games, it would not prove that a draw is the inevitable product of optimal play, just as White's advantage in standard chess does not prove that it can always win. As advanced as machine-learning programs are today, they are not able to mimic perfectly all the moves prescribed by backward induction from every possible endpoint—a draw or a win for one player—in chess.

In section 3, we offer reasons why we think our proposed reversal of two opening moves would tend to equalize the probability that either Black or White can win and, consequently, make the game fairer. Other proposed rule changes that might render chess more balanced or provide other desirable changes (e.g., speed up play) are extensively analyzed in Tomašev, Paquet, Hassabis, and Kramnik (2020). None of this study's rule changes, however, such as the elimination of castling, is as simple as our reversal of two opening moves in chess or as likely to create balance between the two players and force a draw with optimal play.

### III. MAKING CHESS FAIRER

Our argument for changing the order of two opening moves is theoretical: The new order lies between the present one that favors White and a comparable sequence, which we discuss below, that favors Black. The favoritism each player obtains from a sequence, we postulate, is mainly a function of being able to move earlier than its opponent or sometimes later, after observing its move.

Moving earlier gives a player a greater opportunity to "set the stage," whereas moving later enables a player to observe the move of its opponent and respond to it. These two factors are in conflict, creating a tension between moving earlier or later.

There are occasions in chess in which responding to a move can put a player in a more advantageous position than moving first, which is called *Zugzwang* (see https://en.wikipedia.org/wiki/Zugzwang). However, these occasions almost always arise late in a game, when a player's king is in peril; we know of no instances in which Zugzwang can or has happened in the opening moves of chess, which is the kind of change we focus on here.

We assume, consistent with the statistics in section 2, that White has an advantage over Black with the standard chess sequence,

$$\text{WB/WB/WB}…. \quad (1)$$

Black can counter this advantage if it has two moves in a row after White's initial move (WBB), followed by the alternating sequence beginning with W, WB/WB/WB…. The resulting sequence WBB/WB/WB/WB…, can be written as

$$\text{WB/BW/BW/BW}…, \quad (2)$$

where the slashes separate pairs of moves.

Observe that both (1) and (2) start with WB, but the alternation from the 3$^{rd}$ move on of (1) is WB/WB/WB…, whereas that for (2) is BW/BW/BW…. This makes (1) *White favorable*, because W precedes B for every pair from the 3$^{rd}$ move on. By comparison, with (2), B precedes W from the 3$^{rd}$ move on, so this sequence is *Black favorable*.

In effect, giving Black a double move after White's initial move swings the pendulum from favoring White with the standard chess sequence to favoring Black because of the potency of two moves in a row for Black. This raises the question of whether there is an intermediate sequence that creates a balance—between the White favorable sequence of (1) (standard chess) and the Black favorable sequence of (2)—that does not favor either side?

If we add a double move by White (WW) immediately following Black's double move (BB) in the Black favorable sequence of (2), then Black (B) and White each have two moves in a row (BBWW), followed by the alternating sequence

beginning with B on the 6th move (underscored). This gives W/BB/WW/<u>B</u>W/BW/BW…, which can be written as

$$WB/\underline{BW}/WB/WB/WB\ldots, \qquad (3)$$

which we call the *balanced sequence*, or *Balanced Alternation*.

Except for reversal of the 3rd and 4th moves from WB to BW (underscored), the balanced sequence is the standard chess sequence. (Under Balanced Alternation, we assume that a player can check or capture only on the second move of a double move to ensure that the opponent can respond to it.) The balanced sequence is bracketed by the White favorable (standard) sequence of (1) and the Black favorable sequence of (2), making (3) a neutral alterative that favors neither White nor Black.

Notice that there is no difference in the first four moves of (2) and (3), which both start with WB/BW. The difference lies in the alternating sequences, beginning on the 6th move, with the alternation of (2), BW/BW/BW…, favoring Black and the alternation of (3), WB/WB/WB…, favoring White. The alternation of (3) offsets Black's early double move of (2) and provides—with White's later double move of (3)—the balance we seek.

This is not to say that we can prove that the Balanced Alternation of (3), which translates into a switch of the 3rd and 4th moves of standard chess, always leads to a draw with optimal play. But because Balanced Alternation neutralizes (i) the apparent advantage that White derives from (1)—moving first in standard chess—and (ii) the apparent advantage that Black derives from (2) with a single double move, it seems more likely to force a draw with optimal play than either (1) or (2). This does not rule out the possibility that (1), (2) or both, though biased, also force a draw with optimal play.

It is worth pointing out that the first four moves of (1) and (2), WBBW, are identical. They are also the same as those given by the Prouhet-Thue-Morse sequence (see https://en.wikipedia.org/wiki/Thue–Morse_sequence), which can be written as

$$WBBW/BWWB/BWWB/WBBW\ldots, \qquad (4)$$

and is what Brams and Taylor (1999) also call "balanced alternation," as opposed to the "strict alternation" of WB/WB/WB… that is the standard chess sequence. Although the Prouhet-Thue-Morse sequence is arguably fair, it has the disadvantage of allowing more than two double moves, as in the first 16 moves of (4) which contain four double moves.

Another well-known sequence, known as Marseillais chess (see https://en.wikipedia.org/wiki/Marseillais_chess), uses only double moves,

$$WW/BB/WW/BB/WW/BB\ldots. \qquad (5)$$

Like the Prouhet-Thue-Morse sequence, (5) critically alters the strategy and the tactics in chess.

To make a rule change acceptable to chess players, we think it should not have more than two double moves, as does (3). The fact that (3) requires only a reversal of the 3rd and 4th moves from WB to BW makes it even more palatable—it's only a minor revision in the order of play at the beginning of a game, not, as with (4) and (5), throughout. Besides being impartial, (3) seems to be the sequence most likely to force a draw with optimal play, which would make it fair as well.

IV. CHESS ANALYSIS

We next present evidence for the greater fairness of Balanced Alternation by analyzing how moves under it might play out following four well-known first moves by White: (i) 1. e4; (ii) 1. d4; (iii) 1. Nf3; and (iv) 1. c4. From these opening moves, we evaluated the most plausible positions that would be reached under Balanced Alternation by applying some basic chess principles (e.g., winning a free pawn in the opening is generally beneficial) and standard chess reasoning.

To assess the positions that would be reached after the first five moves, W/BB/WW—once Black and White have each made their double moves—we used the Stockfish 13 NNUE chess engine on the Lichess website and on Chessify server. Stockfish recently won the superfinals of the TCEC Season 20 against Leela Chess Zero and is currently considered one of the top chess engines. Although there are tens of thousands of distinct positions that that can occur after the first five moves, we focus only on the four that begin with White's aforementioned first moves and are followed by Black's and White's arguably strongest double moves in response:

**(i) 1. e4:** Black responds with d5 and then captures the e4 pawn by dxe4. White recaptures Black's e4 pawn by first choosing a move such as Nc3, d3, or f3. If White chooses to recapture the pawn with a knight, then the resulting position would be the one given by the following notation: 1. e4 d5 2. dxe4 (B) Nc3 (W) 3. Nxe4, where the only difference with respect to the standard chess notation is that the order of moves under (2) is reversed, as indicated by B and W in parentheses. This brings us to an already "novel" position that is unlikely to occur under a standard chess sequence when White plays 1. e4. Stockfish gives an even "0.0" evaluation to this position. (The depth is 46 in the "cloud," which indicates that the evaluation of this position was already computed and available.) However, if White captures Black's e4 pawn with another piece, then Stockfish gives Black a slight advantage, which would be even greater if White chose not to capture the pawn on e4. Overall, Black is better off under these lines than, for example, under Ruy Lopez (i.e., 1. e4 e5 2. Nf3 Nc6 3. Bb5), where Stockfish's evaluation is +0.2 (depth 56), favoring White by 2/10 of a pawn (positive values indicate a better position for White).

**(ii) 1. d4:** Black responds by playing d5 followed by either e6 or c6, which may lead to standard defenses such as the Queen's Gambit (Declined) and the Slav Defense. In addition, Black can create at least two novelties under Balanced Alternation, capturing White's d4 pawn either with its (a) c-pawn or (b) e-pawn:

—(a) Suppose that Black captures White's d4 pawn by playing c5, followed by cxd4, and White responds by playing c3 followed by cxd4, which leads to the position given by 1. d4 c5 2. cxd4 (B) c3 (W) 3. cxd4. In this position, Stockfish's evaluation is +0.1 (depth 51), which slightly favors White.

—(b) Suppose that Black plays e5 followed by exd4, and White captures the d4 pawn by Qxd4, followed by Nc3, leading to the position given by 1. d4 e5 2. exd4 (B) Qxd4 (W) 3. Nc3. Stockfish's evaluation of this position is +0.1 (depth 40).

Although (a) and (b) slightly favor White (1/10 of a pawn), Black is slightly better off under these lines than the following classical position that arises in the Queen's Gambit: 1. d4 d5 2.c4 e6 3. Nc3, where Stockfish's evaluation is +0.2 (depth 46).

**(iii) 1. Nf3:** Black has solid and well-studied defense options against 1. Nf3, such as d5 and e6, or d5 and c6. White can then create two reasonable novel lines, 1. Nf3 d5 2. e6 (B) e4 (W) 3.exd5 and 1. Nf3 d5 2. e6 (B) e4 (W) 3. e5. The evaluations of both positions are 0.0 (depth 47 and 49, respectively). In addition, the position after 1. Nf3 d5 2. e6 (B) c4 (W) 3. d4 can be reached under both Balanced Alternation and the standard chess; its evaluation is 0.0 (depth 47).

**(iv) 1. c4:** Black plays a standard defense against the English opening. Alternatively, Black may create two novelties by capturing the c4 pawn with either its d-pawn or b-pawn; the latter choice is favored by Stockfish. White's most promising continuation is to capture Black's c4 pawn by playing b3 and bxc4, resulting in the position given by 1. c4 b5 2. bxc4 (B) b3 (W) 3. bxc4. In this position, Stockfish's evaluation toggles between 0.0 and +0.1 as the depth increases, and stabilizes at 0.0 from depth 41 up to and including 47. By contrast, White keeps at least a slight advantage in the English opening under standard chess. (At a lesser depth (up to 30), we have also run Leela Chess Zero evaluations on the positions we present, but we have not obtained any significant differences compared to Stockfish's evaluations.)

Needless to say, our reporting of the evaluations of the novel lines under Balanced Alternation should not be considered as final evidence that those lines are preferable for Black over more established defenses. Our aim in this section has been to illustrate that there are some new plausible lines of defense for Black that seem to create balanced (such as under 1. e4 and 1. c4) or near balanced (e.g., under 1. d4) positions—while ensuring that some of Black's standard defenses are available (e.g., in response to 1. Nf3).

In summary, Black does strictly better under Balanced Alternation than under standard chess for the first moves (i), (ii), and (iv). Only under (iii) are the evaluations the same as those for standard chess. All in all, we believe our analysis shows that Black will be able to do at least as well, and sometimes better, than it does under standard chess.

## V. CONCLUSIONS

The Balanced Alternation of (3) is relatively easy to implement, reversing only the 3$^{rd}$ and 4$^{th}$ moves of standard chess. Our study indicates that Balanced Alternation is more likely to be fairer to Black than (1), and fairer to White than (2), and more likely than either to force a draw with optimal play.

Balanced Alternation makes tournament play more efficient. Presently, because of the bias in favor of White in standard chess, contestants in most tournaments must play White and Black an equal number of times to neutralize this bias.

But this would probably not be necessary with Balanced Alternation. In knockout tournaments, in particular, immediate elimination could occur without the need to play White and Black an equal number of times. Thereby tournaments could accommodate twice as many contestants, each playing as few as one game before elimination. To be sure, physical limitations are not a problem with computer tournaments, but we presume that the attraction of in-person play will resume after the Covid-19 pandemic has subsided.

Without strictly alternating moves from the beginning, there will certainly need to be an adjustment in players' thinking about optimal openings. White, for example, might make a different first move under Balanced Alternation than it would under strict alternation. And Black, able to make two moves in a row (BB) after White's initial move, might use each move for a different purpose—or use both moves to launch a broader attack or defense—as might White with its subsequent double move (WW).

We believe that our reforms are compelling for two reasons: (i) they require only a switch in the order of the 3$^{rd}$ and 4$^{th}$ moves; and (ii) they would not be difficult for chess players to learn and adapt to, after which the familiar alternating moves of WB/WB/WB… occur. Perhaps the main benefit of Balanced Alternation is that it almost surely will make chess fairer, putting Black on a par with White, even if we cannot guarantee that it forces a draw with optimal play.

This is not to say that there may not be other sequences, such as giving only Black a double move, but later than we assumed in (2). What especially appeals to us about (3) is that the reversal occurs early and strikes an even balance between the White favorableness of (1) and the Black favorableness of (2).

Finally, we wish to make clear that we have not proved that Balanced Alternation equalizes the chances of White and Black winning, much less that it forces a draw with optimal play. By giving a boost to Black that puts it on a par with White, Balanced Alternation renders chess fairer than any other reform of which we are aware.

There is a catch, however, that some aficionados of chess worry about—namely, that it will make the game even more drawish than it presently is and hence more boring to watch, especially at the elite level. This is not an unfounded concern. Our view is that new excitement can be injected into the game by giving human players less time to make moves (and so be more prone to make mistakes)—without necessarily going to the extreme of blitz games—that has the advantage of making games shorter and more immediately engaging.